\documentclass[11pt]{article}
\usepackage{moriond,epsfig}

\newcommand{\HyperCP}
           {\mbox{\slshape Hyper}\raisebox{-.2ex}{\slshape C}%
                                 \raisebox{.2ex}{\slshape P}}

\newcommand{\CP}{\mbox{\it CP}}
\newcommand{\ra}{\mbox{~$\rightarrow$}~}

\newcommand{\lm}{\mbox{\it l$^-$}}
\newcommand{\lp}{\mbox{\it l$^+$}}

\newcommand{\elp}{\mbox{$e^+$}}
\newcommand{\elm}{\mbox{$e^-$}}

\newcommand{\gmu}{\mbox{$\mu$}}
\newcommand{\mum}{\mbox{$\mu^-$}}
\newcommand{\mup}{\mbox{$\mu^+$}}

\newcommand{\gpi}{\mbox{$\pi$}}
\newcommand{\pip}{\mbox{$\pi^+$}}
\newcommand{\pim}{\mbox{$\pi^-$}}
\newcommand{\pipm}{\mbox{$\pi^{\pm}$}}

\newcommand{\Kp}{\mbox{$K^+$}}
\newcommand{\Km}{\mbox{$K^{-}$}}
\newcommand{\Kpm}{\mbox{$K^{\pm}$}}

\newcommand{\gp}{\mbox{$p$}}

\newcommand{\gL}{\mbox{$\Lambda$}}

\newcommand{\Xipm}{\mbox{$\Xi^{\pm}$}}

\def\kpee{K^+ \rightarrow \pi^+ e^+ e^-}
\def\kp3pi{K^+ \rightarrow \pi^+ \pi^+ \pi^-}
\def\km3pi{K^- \rightarrow \pi^- \pi^- \pi^+}
\def\kpm3pi{K^{\pm} \rightarrow \pi^{\pm} \pi^+ \pi^-}

\def\kmu{K^{\pm}_{\pi \mu \mu}}

\def\Rkppmm{\Gamma(K^+_{\pi\mu\mu})}
\def\Rkmpmm{\Gamma(K^-_{\pi\mu\mu})}

\def\Delkpm{\Delta(K^{\pm}_{\pi\mu\mu})}
\def\kpmu{K^+_{\pi\mu\mu}}
\def\kmmu{K^-_{\pi\mu\mu}}
\def\kpmmu{K^{\pm}_{\pi\mu\mu}}

\def\kpmpi{K^{\pm}_{{\pi}3}}

\def\kee{K_{{\pi}ee}}
\def\kpee{K^+_{{\pi}ee}}

\def\kpmee{K^{\pm}_{{\pi}ee}}
\def\kmu{K_{\pi\mu\mu}}
\def\kpmu{K^+_{\pi\mu\mu}}
\def\kmmu{K^-_{\pi\mu\mu}}
\def\kpmmu{K^{\pm}_{\pi\mu\mu}}

\def\kpmpi{K^{\pm}_{\pi\pi\pi}}
\def\nokmu{N^{obs}_{\pi\mu\mu}}

\def\nokpi{N^{obs}_{\pi\pi\pi}}

\def\be{\begin{equation}}
\def\ee{\end{equation}}
\def\bea{\begin{eqnarray}}
\def\eea{\end{eqnarray}}

\begin{document}
\vspace*{4cm}
\title{FLAVOR CHANGING KAON DECAYS FROM HYPERCP: \\[0.05in]
       MEASUREMENTS OF THE \Kpm \ra \pipm\mup\mum\ BRANCHING RATIOS}

\author{E.\ Craig Dukes\footnote{Representing the \HyperCP\
collaboration:
        A.~Chan, Y.C.~Chen, C.~Ho, P.K.~Teng,
        {\em (Academia Sinica)};
        W.S.~Choong, G.~Gidal, Y.~Fu, P.~Gu, T.~Jones, K.B.~Luk, B.~Turko,
        P.~Zyla,
        {\em (Berkeley and LBNL)};
        C.~James, J.~Volk,
        {\em (FNAL)};
        J.~Felix,
        {\em (Guanajuato)};
        R.A.~Burnstein, A.~Chakravorty, D.M.~Kaplan, L.M.~Lederman,
        W.~Luebke, D.~Rajaram, H.A.~Rubin, N.~Solomey, Y.~Torun,
        C.G.~White, S.L.~White,
        {\em (IIT, Chicago)};
        N.~Leros, J.P.~Perroud,
        {\em (Lausanne)};
        H.R.~Gustafson, M.J.~Longo, F.~Lopez, H.K.~Park,
        {\em (Michigan)};
        K.~Clark, M.~Jenkins,
        {\em (S.\ Alabama)};
        E.C.~Dukes, C.~Durandet, R.~Godang, T.~Holmstrom, M.~Huang,
        L.~Lu, K.S.~Nelson,
        {\em (Virginia)}.
}}

\address{Physics Department, University of Virginia, \\
Charlottesville, VA 22901, U.S.A.}

\maketitle\abstracts{
The Fermilab \HyperCP\ collaboration is making precision
studies of charged hyperon and kaon decays, as well as
searches for rare and forbidden hyperon and kaon decays.
We report here on measurements of the branching ratios
of the flavor-changing neutral-current decays: 
\Kpm \ra \pipm\mup\mum, and compare our results to theoretical predictions.
This is the first observation of the \Km \ra \pim\mup\mum\ decay.
}

\section{Introduction}

The radiative, flavor-changing neutral-current decays of the type 
\Kpm \ra \pipm\lp\lm\ were originally of considerable interest
because, as tree-level forbidden decays in the standard model, 
they were important systems in which to search for new physics.
The interest in these decays has not waned, despite the lack of any indication
of new physics, but the focus has shifted to
using them as a test-bed for techniques in calculating 
quantum-chromodynamic (QCD) corrections to weak decays.
Expected to proceed via one-loop level electroweak 
box and penguin diagrams, long-distance effects are thought to
play a dominant role in these decays.
Hence chiral perturbation theory (ChPT) has been
employed to calculate both the rates and the form factors of these decays.
An early ChPT calculation which parameterized 
the rates and form factors in terms of a single parameter,
$w_+$,\cite{ecker} has recently been supplemented by a higher-order, 
$\mathcal{O}$$(p^{6})$, ChPT calculation which 
parameterizes them in terms of two variables: $a_+$ and $b_+$.\cite{modin}
This last calculation makes a model-independent prediction that the ratio
$\mathcal{R} = \Gamma(\Kp \ra \pip\mup\mum)/\Gamma(\Kp \ra \pip\elp\elm)$
should be greater than 0.23, a result of rather broad validity since 
it assumes only that the electron and muon channels are described
by the same electromagnetic form factor and that this form
factor does not have singular behavior.
A previous calculation, using a different technique predicts
a similar result.\cite{Singer}

Although first seen in 1975, with the observation
of some forty \Kp \ra \pip\elp\elm\ ($\kpee$) events,\cite{kee1}
only in the last several
years have precision measurements been made of the $\kpee$
branching ratio and form factor in two experiments at 
the Brookhaven National Laboratory (BNL): BNL-777,\cite{kee2} 
and BNL-865.\cite{kee3}
The most recent of these measurements comes from BNL-865.
Using a sample of 10\,300 events they have firmly established
the vector nature of the decay, measured the slope of the form factor
to 9\%, and the branching ratio to 5\%.
The \Kp \ra \pip\mup\mum\ ($\kpmu$) branching ratio
measurement has been more difficult to make.
First observed in 1997 by BNL-787,\cite{kmumu1} they found
a branching ratio about $2\sigma$ less than that expected,
given the prediction of $\mathcal{R}$ mentioned above and 
the BNL-E865 measurement of the $\kpee$ branching ratio.
The $\kpmu$ branching ratio has recently been measured by BNL-865 
with a sample of some 400 events.\cite{kmumu2}
This latter result agrees with the theoretical predictions 
and disagrees with the BNL-787 result by $3.3\sigma$.
Hence it is vital that this discrepancy be resolved
and to determine if indeed the theorists predictions of 
$\mathcal{R}$ are borne out.

\section{The \protect\HyperCP\ Apparatus}

\begin{figure}[bthp]
\centerline{\psfig{figure=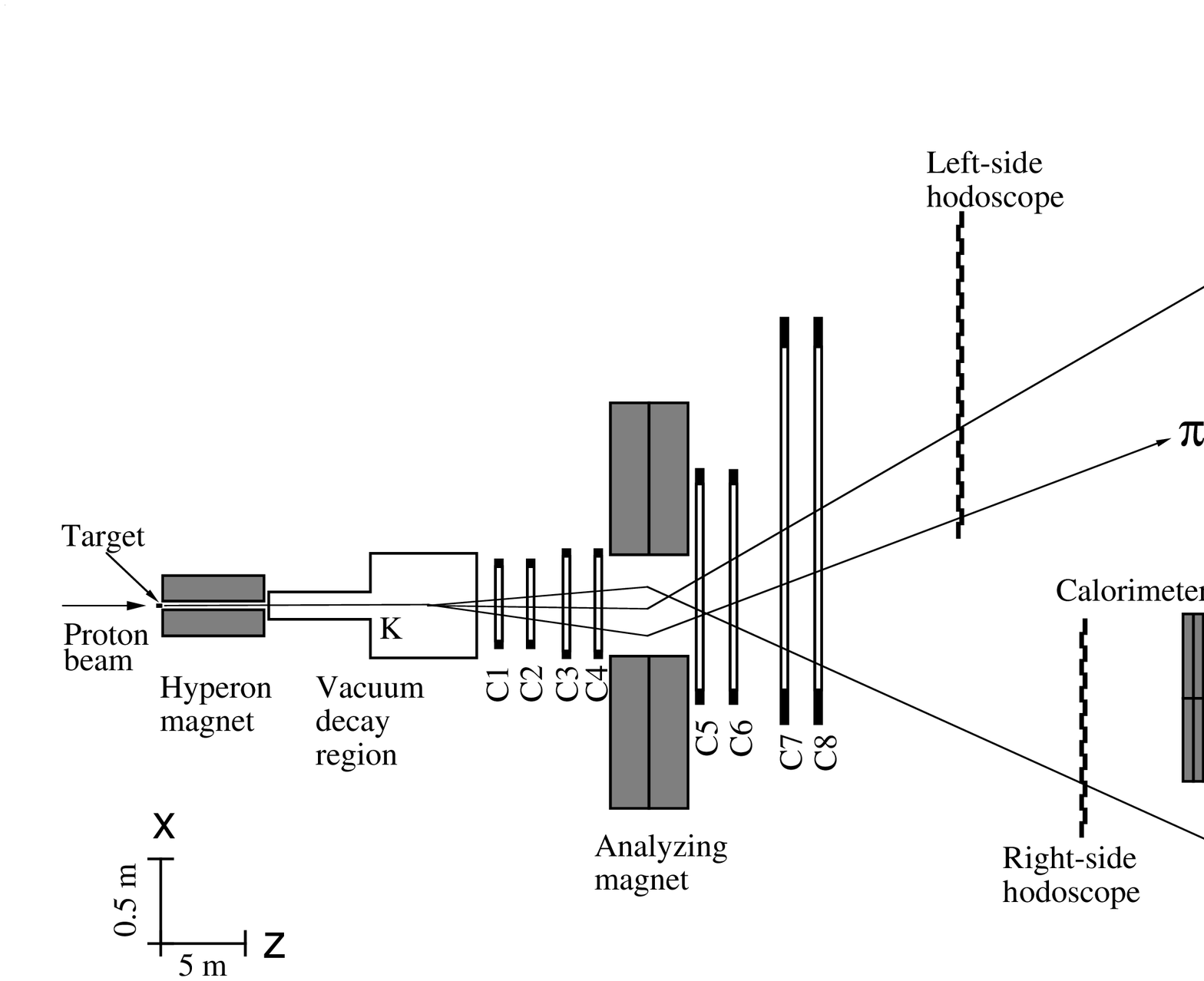,height=60mm}}
\caption{Plan view of the \protect\HyperCP\ spectrometer.
\label{fig:spect_plan}}
\end{figure}

The \HyperCP\ experiment was designed and built primarily
to investigate \CP\ violation in \Xipm \ra \gL\pipm\ hyperon decays.
A plan view of the \HyperCP\ spectrometer is
shown in Fig.~\ref{fig:spect_plan}.
A charged-secondary beam was produced by steering an 800-GeV/$c$ proton
beam onto a $2{\times}2$~mm$^2$-Cu target.
A curved collimator with a 4.88\,$\mu$sr solid angle acceptance 
and embedded in a 6-m long dipole magnet
deflected charged particles, with an average momentum of about 170\,GeV/$c$,
up at a mean angle of 19.5~mrad.
Following a vacuum decay region was a conventional magnetic spectrometer 
employing high-rate, narrow-pitch wire chambers.
Two hodoscopes on either side of the channeled beam
eminating from the collimator exit, called the Left-side and Right-side
hodosopes, were the basis of most of the physics triggers.
A hadronic calorimeter was used for triggering on the proton
from \gL \ra \gp\gpi\ decays, and was not used in the analysis
reported here.
At the rear of the spectrometer was a muon system consisting of two 
identical detectors positioned on either side of
the channeled beam.
The detectors had three proportional-tube stations, each behind
0.81~m of iron absorber, providing $x$ and $y$ measurements.
At the rear of the muon detectors were two hodoscopes, with vertical
and horizontal counters, used to trigger on $\kpmmu$ events.

The polarities of both the Hyperon and Spectrometer magnets
were periodically flipped to change the sign of the 
secondary beam; roughly two-thirds
of the running time was spent on the positive polarity.
A typical primary beam intensity of $6{\times}10^{9}$
protons per second gave a secondary beam rate of 13~MHz
at the exit of the collimator.
Data were taken in two runs, one in 1997, another in 1999,
with a total of 231 billion events recorded on magnetic tape,
of which 555 million were fully-reconstructed 
\Kpm \ra \pipm\pip\pim\ ($\kpmpi$) events.
The analysis reported here is from the smaller 1997 data set
containing 58 billion events.

\section{Analysis}

Rather than measuring absolute branching ratios, which would 
have required
the somewhat difficult task of counting the number of kaons entering the 
apparatus, we measured the $\kpmmu$ branching ratios relative to 
the well-measured (and far larger) $\kpmpi$ branching ratios.
We call $\kpmpi$ the normalization modes and $\kpmmu$ the signal modes.
Since the topologies of both the signal and normalization modes 
were quite similar, 
measuring the branching ratios by this means minimized systematic errors.

The triggers used for both signal and normalization modes were 
both quite simple.  
For the normalization modes the triggers required a
coincidence of signals in both the Left-side and Right-side hodoscopes.
The triggers for the signal modes required this same coincidence along with
a coincidence of charged particles in the Left-side and Right-side 
muon hodoscopes.

The branching ratios for $\kpmmu$ were determined using the following formula:
\begin{eqnarray}
  B(\Kpm \ra \pipm\mup\mum)=\frac{1}{200}\frac{N_{\pi\mu\mu}^{obs}}
                                              {N_{\pi\pi\pi}^{obs}}
          \frac{A_{\pi\pi\pi}}{A_{\pi\mu\mu}}
          \frac{\epsilon_{\pi\pi\pi}^{trig}}{\epsilon_{\pi\mu\mu}^{trig}}
          \frac{\epsilon_{\pi\pi\pi}^{sel}}{\epsilon_{\pi\mu\mu}^{sel}}
          \frac{ B(\Kpm \ra \pipm\pip\pim) }
               {\epsilon^{\mu^+ \mu^-}},
\label{eq:bratio}
\end{eqnarray}
where: $N^{obs}$ are the numbers of observed events,
$A$ are the geometric acceptances of the triggers,
$\epsilon^{sel}$ are the event-selection efficiencies,
and $\epsilon^{trig}$ are the trigger efficiencies;
for signal ($\kpmmu$) and normalization ($\kpmpi$) modes.
The number 200 is the prescale factor for the normalization trigger
and $\epsilon^{\mu^+ \mu^-}$ is the muon identification efficiency.
The values of the parameters used in Eq.~\ref{eq:bratio} are given
in Table~\ref{tab:param} below.  
Note the generally large values of the efficiencies and acceptances.
The following discussion explains
how these values were obtained.
\begin{table}[htpb]
\caption{Parameters used to calculate the branching ratios.
\label{tab:param}}
\mbox{}\\[0.3mm]
\centering
\begin{tabular}{lcc}
\hline \hline
\multicolumn{1}{c}{Item}       & $+$ Polarity      & $-$ Polarity \\
\hline
$\nokmu$                       & $65.3\pm{8.2}$    & $35.2\pm{6.6}$ \\
$\nokpi$ $({\times}10^5)$      & $4.446 \pm 0.010$ & $2.318 \pm 0.008$ \\
$A_{\pi\mu\mu}$                & 94.4\%            & 94.2\% \\
$A_{\pi\pi\pi}$                & 47.5\%            & 47.7\% \\
$\epsilon^{sel}_{\pi\mu\mu}$   & 80.3\%            & 78.3\% \\
$\epsilon^{sel}_{\pi\pi\pi}$   & 77.9\%            & 76.0\% \\
$\epsilon^{trig}_{\pi\mu\mu}/\epsilon^{trig}_{\pi\pi\pi}$
                               & \multicolumn{2}{c}{$(86.6 \pm 2.7)\%$}  \\
$\epsilon_{\mu^+\mu^-}$        & \multicolumn{2}{c}{$(93.8 \pm 0.3)\%$}  \\
$B(\Kpm \ra \pipm\pip\pim)$    & \multicolumn{2}{c}{$(5.59{\pm}0.05)\%$} \\
\hline \hline
\end{tabular}
\end{table}
\begin{figure}[htb]
\begin{minipage}[t]{80mm}
\centerline{\psfig{figure=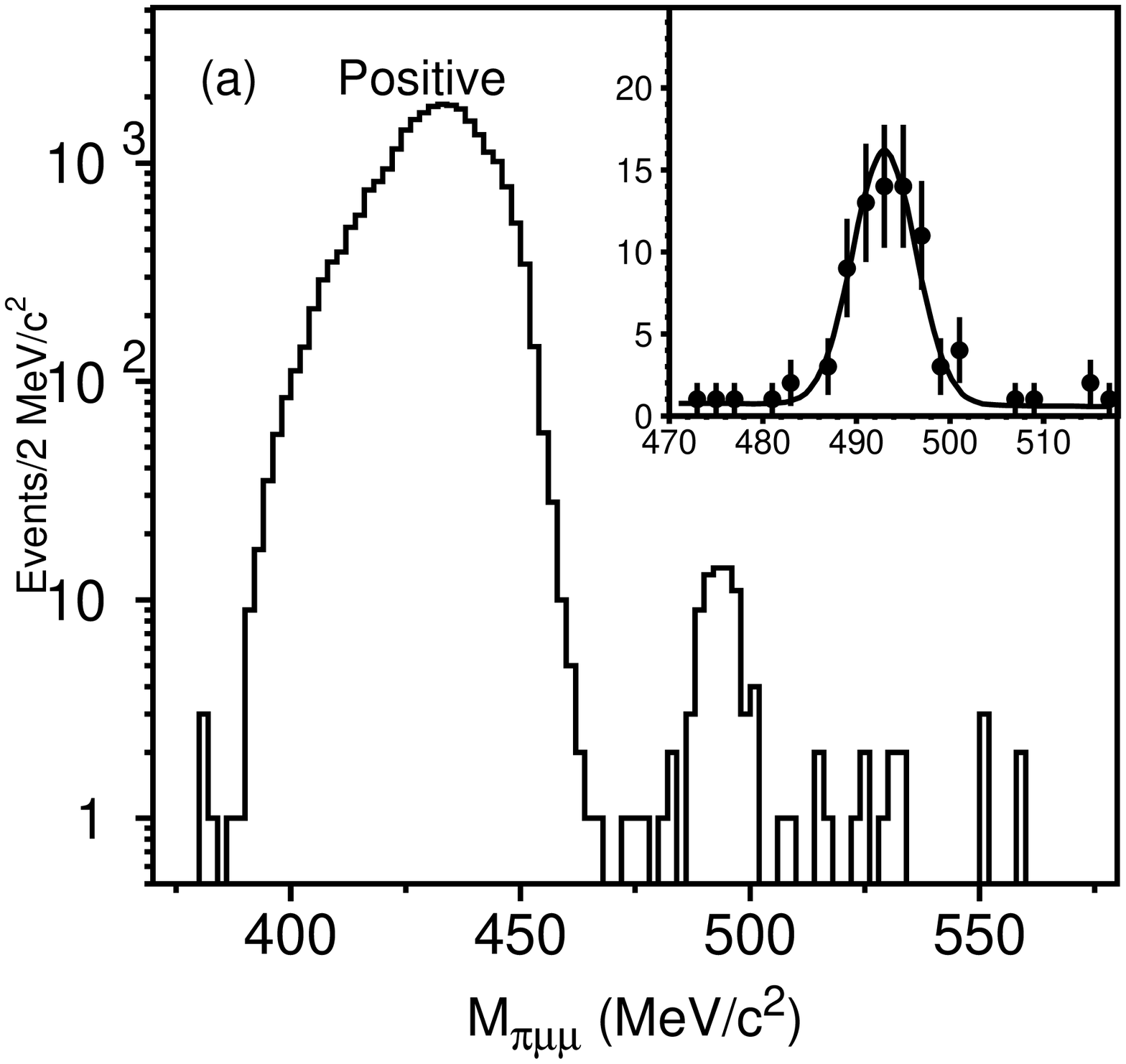,height=65mm}}
\end{minipage}
\hfill
\begin{minipage}[t]{80mm}
\centerline{\psfig{figure=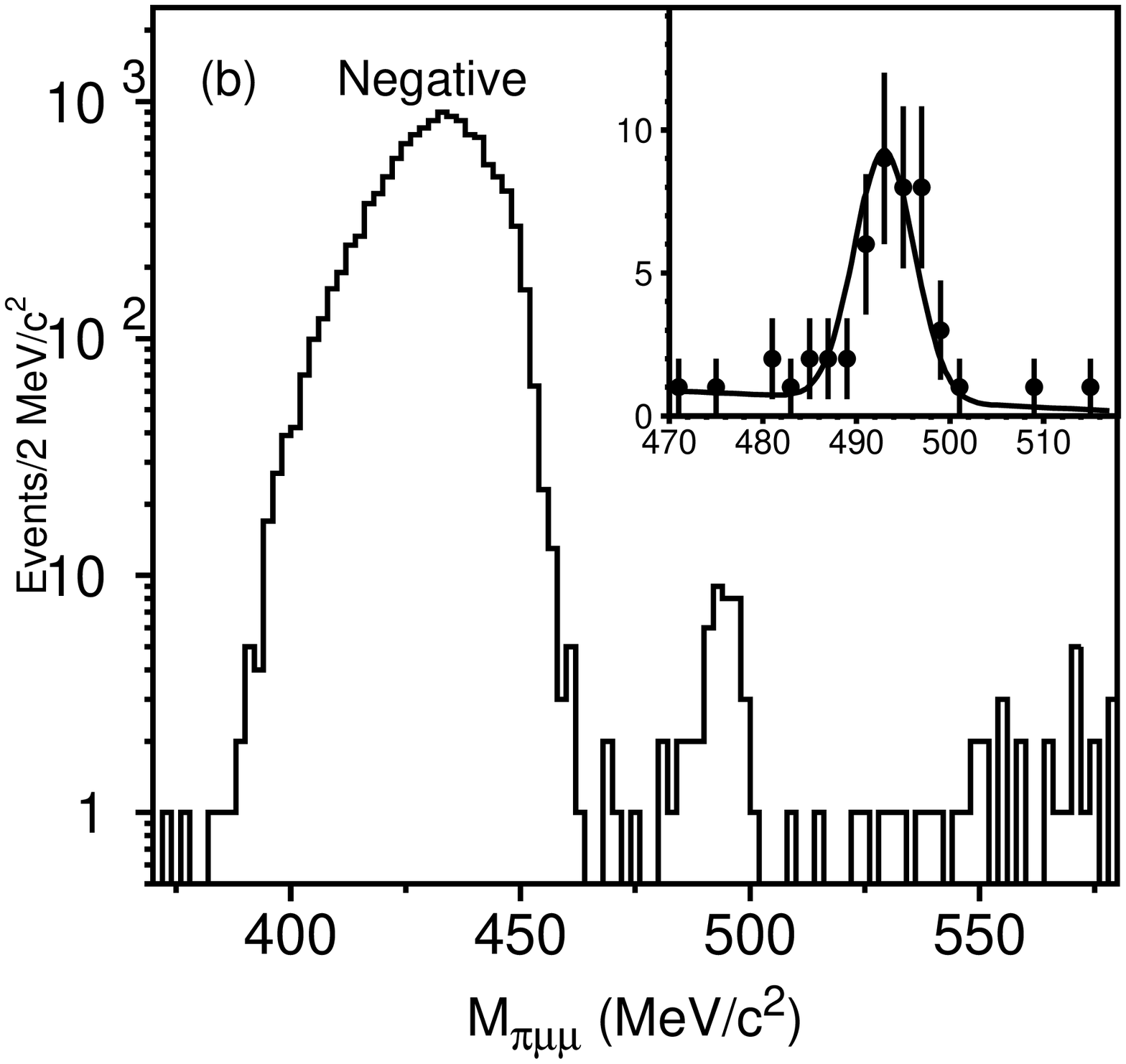,height=65mm}}
\end{minipage}
\caption{The \gpi\gmu\gmu\ invariant mass for positive and 
negative beam data after cuts.  The large peak below the
$\Kpm$ mass is from $\kpmpi$ events with either in-flight
decays of the pions or pion punchthrough.
The insets show the signal region used to establish the number of events.
\hfill
\label{fig:signal_evts}}
\end{figure}

\begin{figure}[htb]
\begin{minipage}[t]{60mm}
\centerline{\psfig{figure=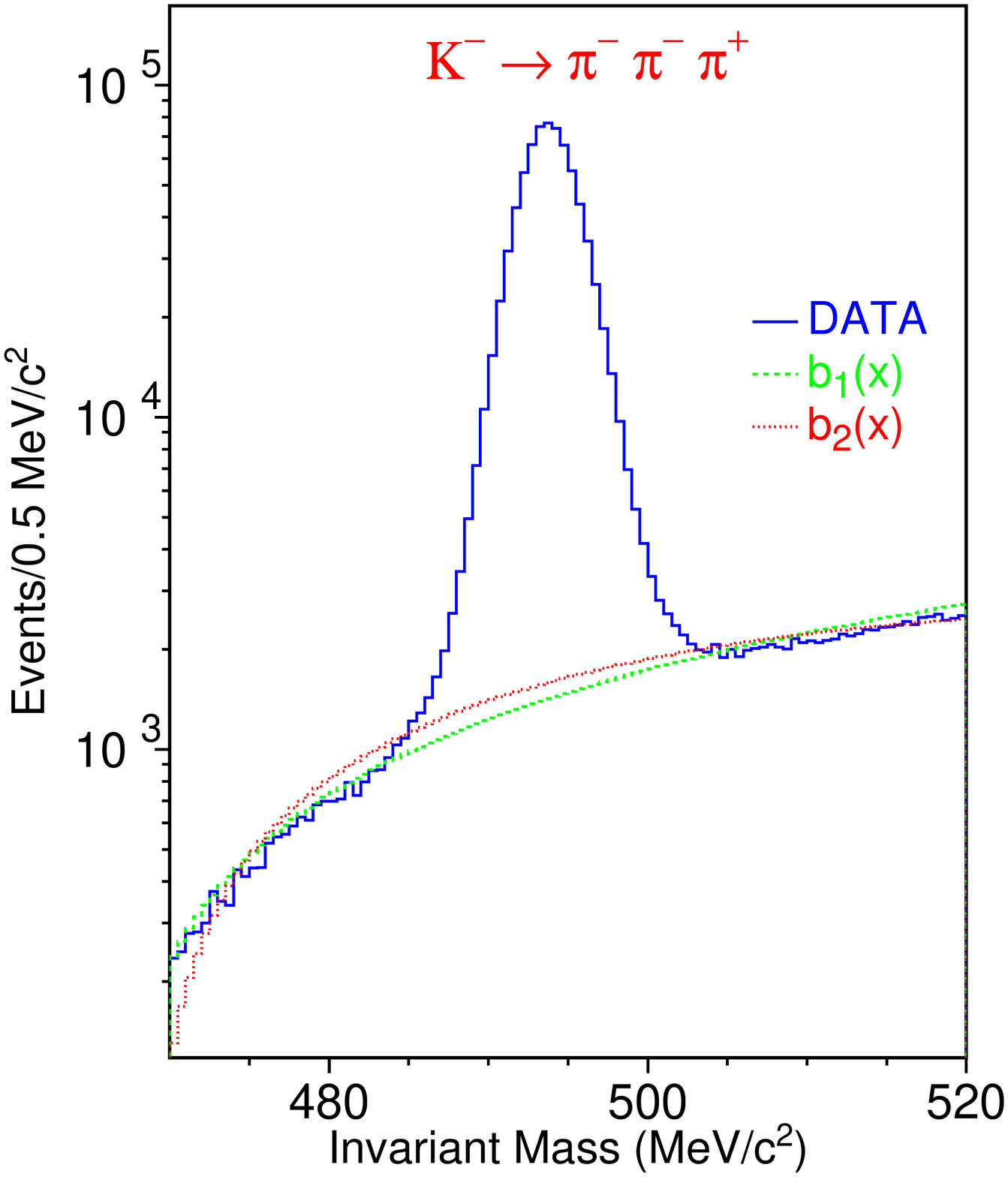,height=80.5mm}}
\caption{The $\pim\pip\pim$ invariant mass.
The dashed and dotted curves, b$_1$ and b$_2$,
are respectively linear and quadratic fits to the
background in the side-bands.\hfill
\label{fig:norm_evts}}
\end{minipage}
\hfill
\begin{minipage}[t]{90mm}
\centerline{\psfig{figure=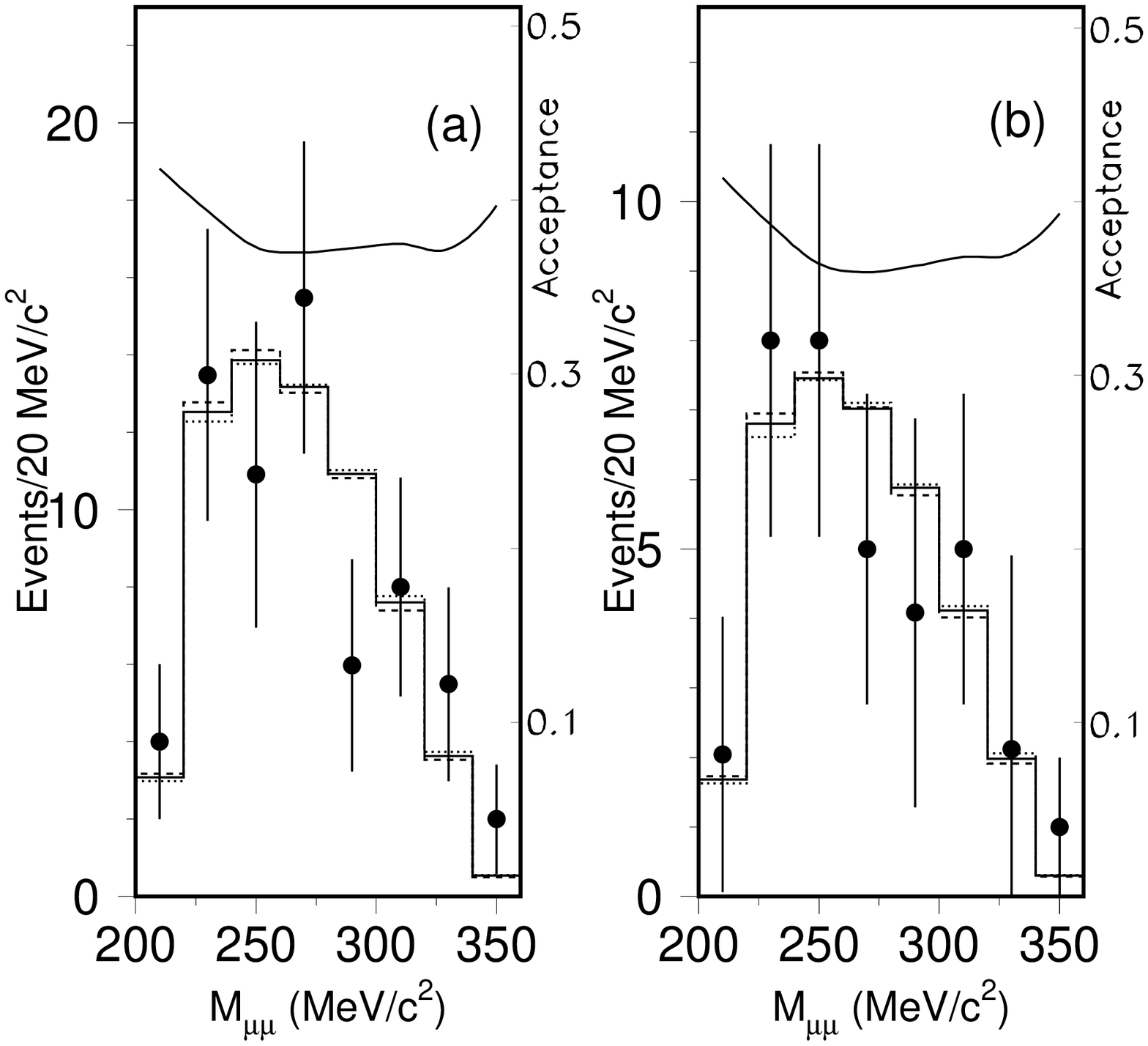,height=80mm}}
\caption{The dimuon invariant mass for (a) positive- and
(b) negative-beam data after cuts.  The solid
lines are the Monte Carlo predictions using the 
measured values of the slope parameters.  The dashed and
dotted lines represent Monte Carlo predictions with
${\pm}3\sigma$ variations, respectively, from the measured value
of the slope parameter.  The curves show the acceptance. \hfill
\label{fig:dimuon_mass}}
\end{minipage}
\end{figure}

Data for both signal and normalization modes were selected in a
similar manner.  A three-track topology --- two on the left side
of the spectrometer and one on the right --- was required in both cases,
with the signal tracks having the additional requirement of a
muon track in both the Left- and Right-side muon detectors.
A muon track was defined as a track stub in a muon detector which
included at least two of three hits in both $x$ and $y$ views
in the muon proportional tubes with corresponding in-time hits
in the muon hodoscopes.
Two of the three pion tracks for the normalization modes
were required to point to the fiducial areas of the muon detectors.
A good fit to a common vertex was required of the three tracks.
This was done by first fitting the tracks to a common vertex and
requiring that the average transverse separation of the three tracks 
at that point be less than 2~mm and the $\chi^2/$d.o.f of the 
vertex-constrained fit be less than 2.5.
This cut out most of the background from hyperon decays and
uncorrelated tracks.
Other rather loose cuts were applied to the parent: 
on its momentum, target point-back, and  decay vertex.
The invariant mass of the signal and normalization modes after these
cuts is shown in Fig.~\ref{fig:signal_evts}.
The number of signal events, as determined by maximum-likelihood
fits in the region 470 MeV/$c^2$ $<$ $M_{\pi \mu \mu}$ $<$ 520 MeV/$c^2$,
is given in Table~\ref{tab:param}.
The invariant mass of normalization data sample is shown in
Fig.~\ref{fig:norm_evts}.

From the combined positive and negative data samples
we measured the relative trigger efficiency,
$\epsilon^{trig}_{\pi\mu\mu}/\epsilon^{trig}_{\pi\pi\pi}$,
and the dimuon selection efficiency: $\epsilon_{\mu^+\mu^-}$,
given in Table~\ref{tab:param}.

From Monte Carlo we determined the acceptances 
and the event selection efficiencies for signal
and normalization modes.  
The Particle Data Group (PDG) parameters describing the 
$\kpmpi$ decays were used.\cite{pdg}
The $\kpmmu$ decay parameters are not as well known.
In the Monte Carlo simulation we assumed a vector interaction, 
as has been demonstrated in $\kpee$ decays by the BNL-865 experiment,\cite{kee3}
and is consistent, although not nearly as well established,
with their $\kpmu$ measurement.\cite{kmumu1}
For the slope of the form factor which describes 
the dilepton invariant mass spectrum, $M_{ll}$, for the $\kpmmu$ decays
we averaged the values from the BNL-777 $\kpee$
and BNL-865 $\kpee$ and $\kpmu$ measurements to obtain
$\delta=2.08 \pm 0.12$:  the form factor being given by
$\phi(M_{ll}^2)=\phi_{0}(1 + \delta M_{ll}^2/m_{K^+}^2)$,
where $\delta$ and $\phi_{0}$ are a slope parameter
and a constant, respectively, and $m_{K^+}$ is the mass of the charged kaon.
It is important to note, as can be seen in Fig.~\ref{fig:dimuon_mass},
that our acceptance in the dimuon invariant mass was
quite uniform and hence we were not sensitive to even large
deviations from the average value of the slope parameter $\delta$.
The geometric acceptances are given in Table~\ref{tab:param}.
The event selection efficiencies, also measured using Monte Carlo,
are quite high, and are also given in Table~\ref{tab:param}.

A word on errors.  
Our measurements were dominated by the statistical uncertainties in 
determining the number of $\kpmmu$ events: giving a 
10\% error for the combined branching ratio measurement.
Sources of systematic error were carefully studied,
including: targetting variations, magnetic field variations,
the difference between data and Monte Carlo, uncertainties
in the parameters describing the $\kpmmu$ and $\kpmpi$ decays,
and uncertainties in the trigger and muon identification efficiencies.
The total systematic errors were found to be less than 
half the statistical errors.
They were dominated by two sources roughly equal in magnitude:
the error in the background under the signal peak (3.6\%) 
and the error in the determination of the relative trigger efficiencies (3.1\%).

\section{Results}

\begin{figure}
\begin{minipage}[t]{75mm}
\centerline{\psfig{figure=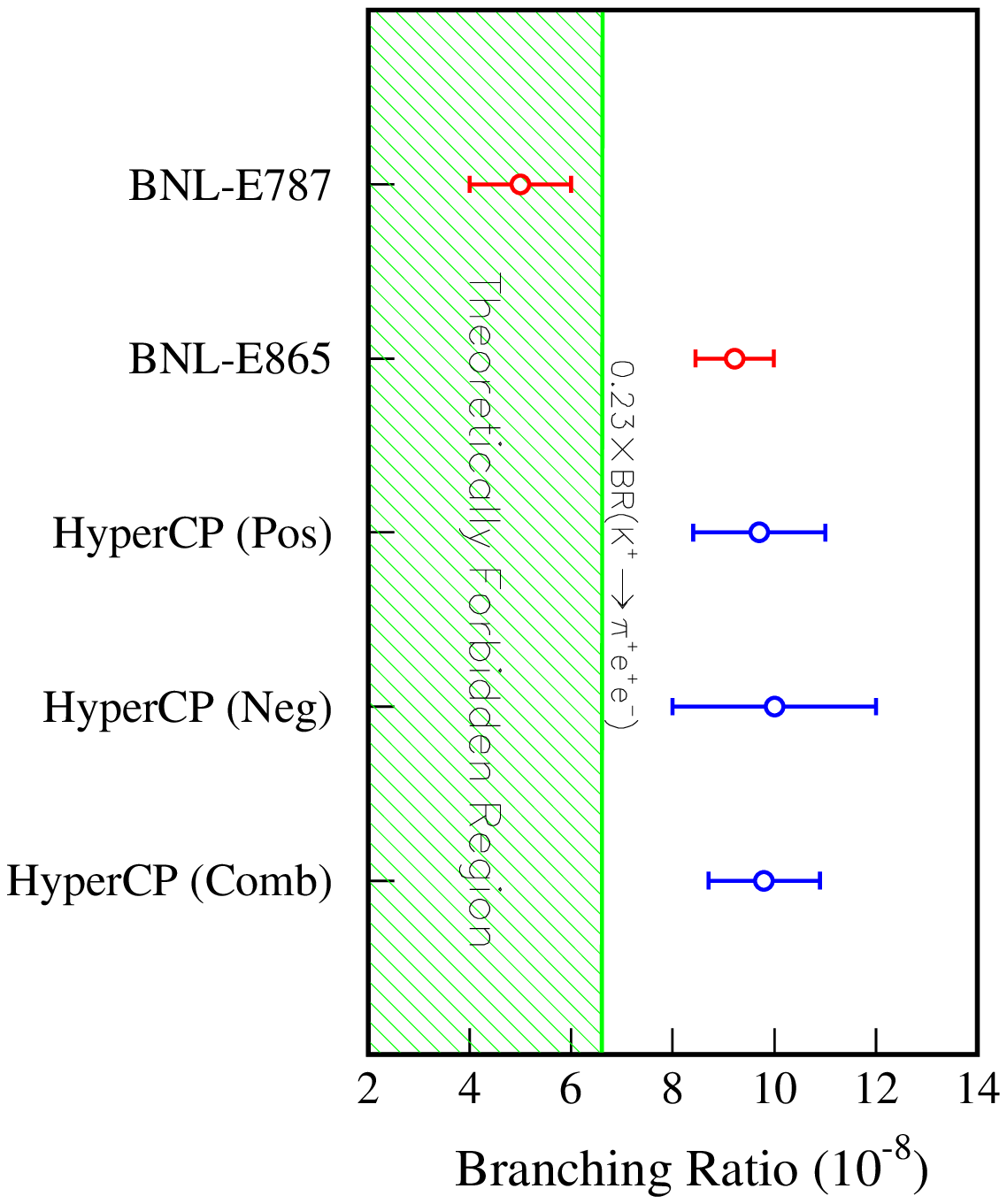,height=85mm}}
\caption{Experimental measurements of the $\kpmmu$ branching ratios.
The two BNL points refer to measurements of $B(\kpmu)$.\hfill
\label{fig:exp_results}}
\end{minipage}
\hfill
\begin{minipage}[t]{75mm}
\centerline{\psfig{figure=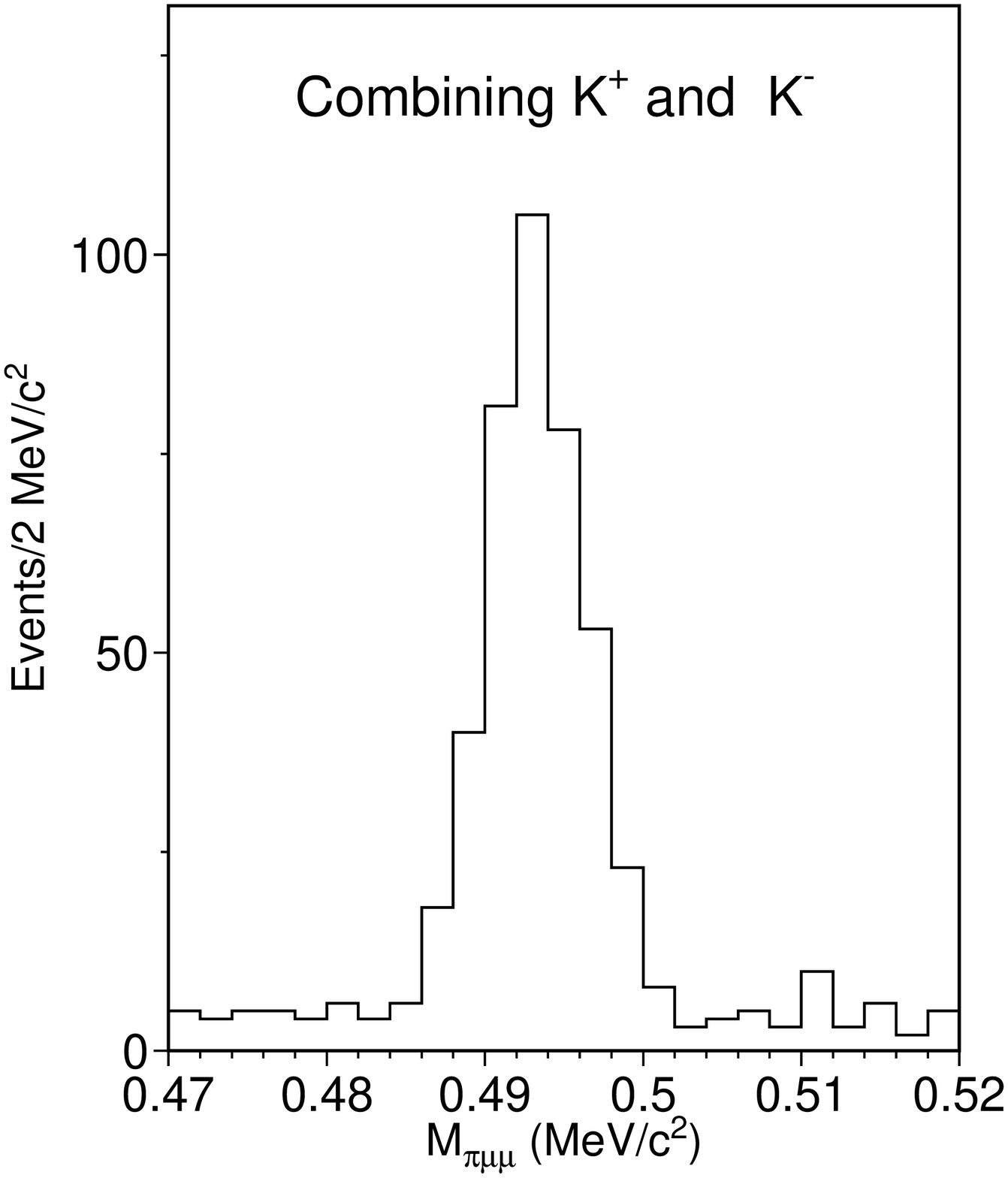,height=85mm}}
\mbox{}\\[-3.0in]
\hspace*{1.0in}\rotatebox{45}{\Huge {Preliminary!}}
\mbox{}\\[1.25in]
\caption{The \gpi\gmu\gmu\ invariant mass from the 1999 \protect\HyperCP\
data set.\hfill
\label{fig:prelim_mass}}
\end{minipage}
\end{figure}

What do we find for the branching ratios?
Using Eq.~\ref{eq:bratio} we get:
$B(\kpmu) = (9.7\pm 1.2\pm 0.4) \times 10^{-8}$
and $B(\kmmu)$ = ($10.0\pm 1.9\pm 0.7) \times 10^{-8}$,
where the first and second errors are statistical and systematic, respectively.
The two results are in agreement with each other, indicating
no evidence of \CP\ violation:
\begin{eqnarray*}
\Delkpm  =  \frac{\Rkppmm - \Rkmpmm}{\Rkppmm + \Rkmpmm}
          = -0.02\pm 0.11{\rm (stat)} \pm 0.04{\rm (syst)}.
\end{eqnarray*}
No surprise here:  theoretical expectations for the \CP\ asymmetry
are orders of magnitude smaller.\cite{cpest}
Assuming \CP\ symmetry is good, we can combine the two
measurements to give:
\begin{eqnarray*}
B(\kpmmu) = [9.8{\pm}1.0{\rm (stat)}{\pm}0.5{\rm (syst)}]{\times}10^{-8}.
\end{eqnarray*}
This result is consistent with the BNL-865 measurement,
but $3.2\sigma$ higher than the BNL-787 result
(see Fig.~\ref{fig:exp_results}).
It also, when combined with the PDG value for $B(\kpee)$,
is consistent with the theoretical predications of 
$\mathcal{R} = \Gamma(\kmu)/\Gamma(\kee)$.

The \HyperCP\ collaboration is analyzing the larger 1999
data set, which should give about a factor of four increase in
statistics, and which has quite different systematics:
a different muon detector, Monte Carlo, and tracking program.
Figure~\ref{fig:prelim_mass} shows the \gpi\gmu\gmu\ mass peak from
a preliminary analysis of the combined $\kpmu$ and $\kmmu$ 1999
data sets.
By the end of the decade, new experiments such as CKM at Fermilab,
should produce data sets for both the $\kpmee$ and $\kpmmu$ modes
that are on the order of 100,000 events, which will allow precision
tests of ChPT predictions to be made.

We finally note that these results have been recently published
in Physical Review Letters.\cite{hypercp}

\section*{Acknowledgments}
We thank the organizers for an interesting and stimulating
conference.
We are indebted to the staffs of Fermilab and
the participating institutions for their vital contributions.
This work was supported by the U.S. Department of Energy and
the National Science Council of Taiwan, R.O.C.
K.B. Luk was partially supported by the Miller Institute.

\end{document}